\documentclass[fleqn,12pt,twoside]{article}
\usepackage{espcrc1}

\usepackage{graphicx}

\newcommand{\sqrtsNN}{\mbox{$\sqrt{\mathrm{s}_{_{\mathrm{NN}}}}$}}
\newcommand{\alam}{\overline{\Lambda}}
\newcommand{\lam}{\Lambda}
\newcommand{\axi}{\overline{\Xi}^+}
\newcommand{\xim}{\Xi^-}
\newcommand{\pim}{\pi^-}

\newcommand{\omm}{\Omega^-}
\newcommand{\aom}{\overline{\Omega}^+}

\title{$\xim$ and $\axi$ baryon production in Au+Au collisions at $\sqrtsNN=130$ GeV}

\author{Javier Castillo\address{SUBATECH, Nantes, France} for the STAR 
collaboration\footnote{For the full author list and acknowledgments, see Appendix ``Collaborations'' of this volume.}
}
 
\begin{document}

\maketitle

\begin{abstract}

We report preliminary results on the centrality dependence of the 
$\xim$ and $\axi$ production at mid-rapidity in $\sqrtsNN=130$ GeV 
Au+Au collisions from the STAR experiment. For the most central data
the obtained yields suggest a saturation of strangeness production per
produced hadron. The calculated inverse slope parameter may indicate an earlier 
freeze-out of these particles.
\end{abstract}




\section{Introduction}

The study of multi-strange baryons in ultra-relativistic heavy ion
collisions was first motivated by the prediction of an enhancement 
of strange $q\overline{q}$ pair production due to the formation of 
a quark gluon plasma (QGP) compared to that found from a hadron gas
state \cite{MR82}. The enhancement is thought to be more pronounced 
for multi-strange baryons and anti-baryons since their production is strongly
suppressed in hadronic interactions due to their high mass
thresholds \cite{R91}. Also, by comparing the multi-strange baryon yields 
and ratios to other particles with those predicted from particle 
production models, one can test whether the system has reached 
chemical equilibrium or not. The study of transverse momentum spectra 
will allow us to test the thermal equilibration of the system. 

The Relativistic Heavy Ion Collider (RHIC) allows for the study of
nuclear matter in extreme conditions such as in the case of the
theoretical QGP. We present here preliminary measurements of
multi-strange baryon production at mid-rapidity as a function of 
the collision centrality in $\sqrtsNN=130$ GeV Au+Au collisions from the
STAR experiment.

\section{Experimental setup and data analysis}

The data presented here were taken with the STAR detector during
the year 2000 Au+Au at $\sqrtsNN=130$ GeV run. The experimental
setup for this period consisted mainly of a large cylindrical Time
Projection Chamber (TPC) used for charged particle tracking and
identification. A Central Trigger Barrel (CTB) and two Zero Degree
Calorimeters (ZDC) were used for triggering. More details of the
experimental setup can be found in \cite{STARlam}. 
The data were divided into three centrality classes corresponding to $0-10 \%$,
$10-25 \%$ and $25-75 \%$ of the total hadronic cross section as described 
in \cite{STARflow}.

The multi-strange $\xim$ particles were reconstructed via their
decay topologies, $\xim \rightarrow \lam \pi^-$ followed by $\lam
\rightarrow p \pi^-$ with a branching ratio of $100\%$ and $64\%$
respectively. The reconstruction of $\xim$ particles was handled by
first reconstructing $\lam$ candidates as described in 
\cite{STARlam}. These $\lam$ candidates were
then extrapolated backwards and combined with negatively charged
tracks to determine $\xim$ candidates, and similarly for the
charge conjugate $\axi$ decay. Particle identification from energy
loss of charged tracks in the TPC as well as simple geometric and
kinematic cuts were applied at both steps to reduce the large
combinatorial background inherent to such a reconstruction
process. For $\xim$ and $\axi$ the raw yields were extracted from the
invariant mass distribution by counting the entries within
$\pm15$~MeV/c$^2$ about the expected mass and then subtracting the
background. The background under the peak was estimated by sampling
two regions on either side.

\section{Results and Discussion}

\begin{figure}
\begin{minipage}[t]{0.52\textwidth}
\includegraphics[width=\textwidth,height=0.8\textwidth]{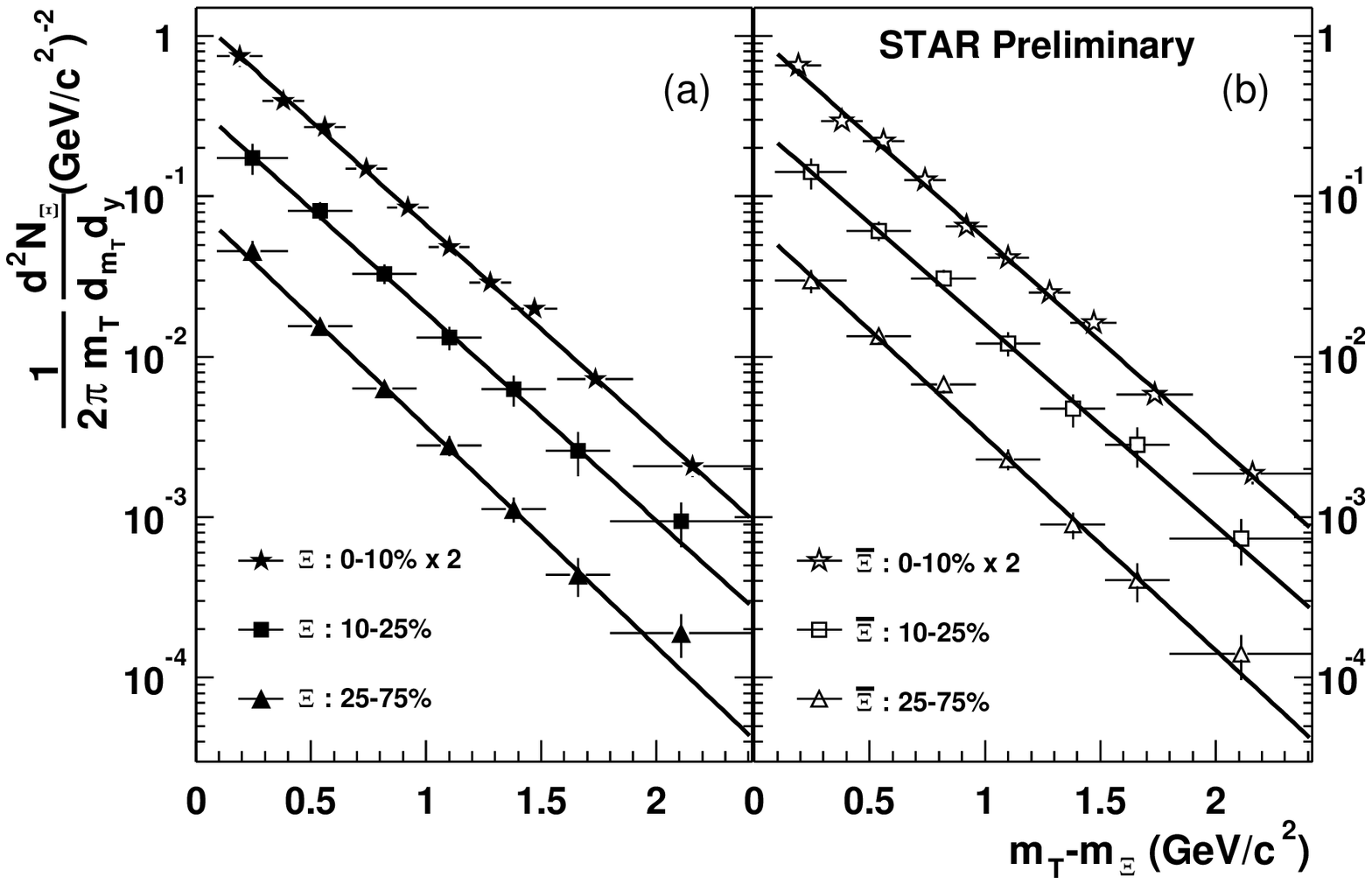}
\caption{Transverse mass spectra of (a) $\xim$ and (b) $\axi$ as a function of 
collision centrality. Lines are exponential fits to the data.} \label{corrXiMt}
\end{minipage}\hfill
\begin{minipage}[t]{0.45\textwidth}
\includegraphics[width=\textwidth,height=0.9\textwidth]{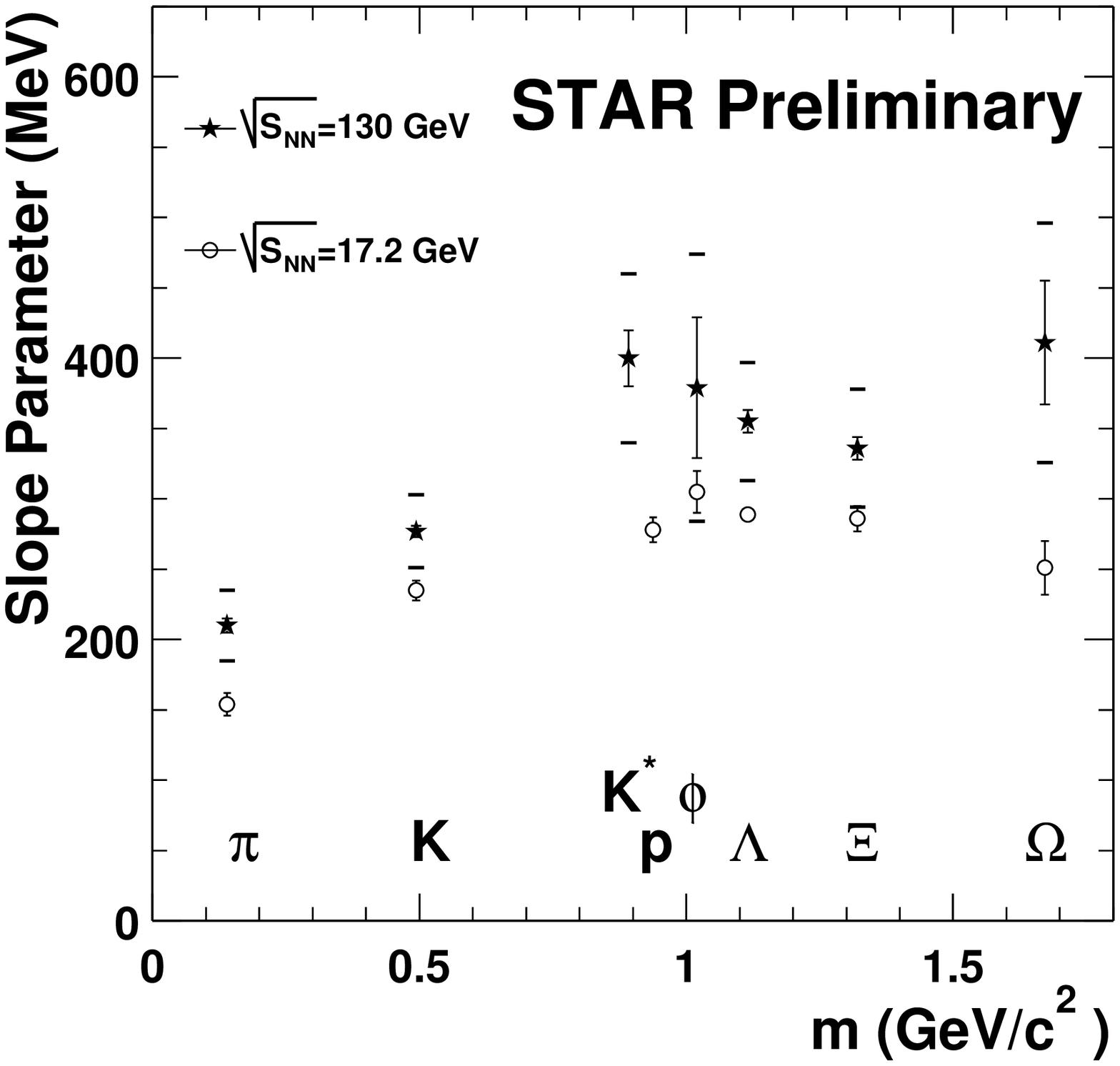}
\caption{Inverse slope parameter as a function of particle mass for (circles) SPS 
and (stars) RHIC central collisions.} \label{invSlopeVsMass}
\end{minipage}\hfill
\end{figure}

The invariant mass distribution of $\xim$($\axi$) candidates was
then histogramed in transverse mass $m_T=\sqrt{p_T^2+m_0^2}$ bins
and the signal extracted in each bin as described above. Each
$m_T$ bin was then corrected for detector acceptance and
reconstruction efficiency by the Monte Carlo technique where
simulated $\xim$ particles were embedded into real events. The
data cover $|y|<0.75$, simulations showed that the acceptance and
efficiency for $\Xi$ is constant, in $y$, over this range.

\subsection{Centrality dependence of invariant yields and slope parameter}

Each of the obtained preliminary $m_T$ spectra for $\xim$ and $\axi$ for 
the three centrality classes used in this analysis was then fit to
an exponential function ($\propto \frac{dN}{dy}e^{-(m_T-m_0)/T}$) to
determine simultaneously the inverse slope, $T$, and the rapidity density,
$\frac{dN}{dy}$, integrated over all $m_T$. The results are summarized in 
table~\ref{fitParXi}. We observe that the slope parameter is, within error bars, 
the same for $\xim$ and $\axi$ and is not strongly dependent on centrality. 
Also the $\frac{dN}{dy}$ for $\xim$ and $\axi$ is found to increase linearly with 
$dN_{h^{-}}/d\eta$. Thus, no noticeable change of the production 
mechanism for $\Xi$ baryons is observed in the covered centrality region.
Systematic errors are estimated to be $\sim 20\%$. The feeddown from weak decay 
of $\omm$($\aom$) on $\xim$($\axi$) is estimated to be less than $2\%$ and thus 
it has been neglected for this analysis. The $\xim$ and $\axi$ particle yields 
in the measured $m_T$ region correspond to $\sim 75\%$ of the total yield. 

\begin{table}
\center
\begin{tabular}{|c|c|c|c|c|}
  \hline
  & Centrality & $0-10\%$ & $10-25\%$ & $25-75\%$ \\ \hline
  $\xim$ & $dN/dy$ & $2.30\pm0.09$ & $1.29\pm0.12$ & $0.28\pm0.02$ \\
  & $T$ (MeV) & $335\pm6$ & $336\pm18$ & $318\pm14$ \\ \hline
  $\axi$ & $dN/dy$ & $1.89\pm0.08$ & $1.08\pm0.11$ & $0.24\pm0.2$ \\
  & $T$ (MeV) & $338\pm7$ & $344\pm18$ & $325\pm14$ \\ \hline
\end{tabular}
\caption{Mid-rapidity fit parameters from exponential fit to $\xim$ and $\axi$ $m_T$ spectra.}
\label{fitParXi}
\end{table}

\subsection{Systematics of the inverse slope parameter}

Figure~\ref{invSlopeVsMass} shows the mid-rapidity inverse slope
parameters as a function of particle mass, for central Pb+Pb
collisions at $\sqrtsNN=$17.2 GeV
\cite{WA97slopes,NA44slopes,NA49phi} and preliminary results from
Au+Au collisions at $\sqrtsNN=$130 GeV
\cite{STARlam,STARovQM01,STARphi,STARomQM02}. Note that both, statistical errors and
systematic errors (horizontal lines) are reported on STAR points, while SPS
points only include statistical errors. We observe the following :
the inverse slope parameters are higher in RHIC collisions than in SPS
collisions and the strange baryons seem to deviate from the collective
flow pattern driven by lighter particles at both energies. 
The relatively low inverse slope parameter found for the $\Xi$ baryons may 
be due to the functional form and the $m_T$ range used for the fit or it 
could indicate, as proposed in \cite{VanHecke98}, that they freeze-out earlier 
during the system's evolution.

\subsection{Non-identical particle ratios}

\begin{figure}
\center
\includegraphics[width=0.7\textwidth]{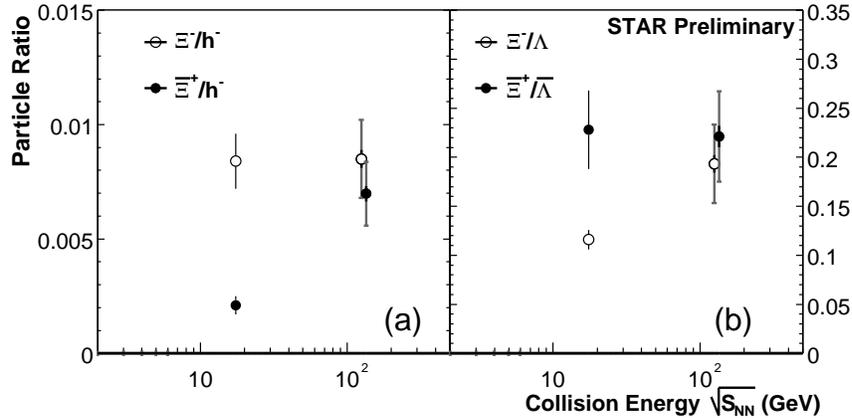}
\caption{Mid-rapidity (a) $\Xi / h^-$ ratios and (b) $\Xi /
\Lambda$ ratios as a function of beam energy. Systematic errors to 
STAR points are shown in gray.}
\label{nonIdenRatios}
\end{figure}

Figure~\ref{nonIdenRatios} shows the mid-rapidity ratios (a) $\Xi
/ h^-$ and (b) $\Xi / \Lambda$ from heavy ion collisions at 
SPS \cite{WA97yields}, plus preliminary RHIC results.  
The $\axi / h^-$ ratio increases with beam energy indicating a higher exitation 
energy from the collision reached at RHIC energies compared to SPS energies. 
On the other hand, the $\xim / h^-$ ratio remains constant from SPS
to RHIC energies due to the competing effect of the droping net-baryon density. 
Also from SPS to RHIC energies, the $\xim / \lam$ ratio increases due again to 
the drop of the net-baryon density which causes a smaller fractional rise on $\lam$ 
than on $\xim$, while the $\axi / \alam$ is constant. The behaviour 
of $\axi / \alam$ suggests a saturation of the strangeness production 
from SPS to RHIC. Note that $\xim / \lam$ and $\axi / \alam$ ratios 
have been corrected for feeddown from weak decays of $\Xi$ on $\Lambda$. 
For SPS values the quoted \cite{WA97slopes} upper limits of $5\%$ for $\lam$ 
and $10\%$ for $\alam$ were used.

Particle ratios between dissimilar species can be used to test particle
production models. Using preliminary STAR data we have obtained 
$\xim / \pim = 0.0088\pm0.0004$. 
ALCOR, a quark coalescence model, overestimates 
the $\xim / \pim$ ratio by almost a factor of two \cite{LevaiSQM01}. 
A statistical model assuming local chemical equilibrium and strangeness 
neutrality \cite{PBM01} is, within systematic errors, in agreement 
with the experimental value of $\xim / \pim$ ratio. Finally the authors of 
\cite{Rafelski02} need to assume chemical non-equilibrium 
in their QGP model to better address the $\xim / \lam$ ratio. 

\section{Conclusion}

In summary, the $\xim$ and $\axi$ invariant yields at mid rapidity increase linearly with the 
number of produced negative hadrons. The production of strangeness per produced
hadron seems to saturate from SPS to RHIC energies, which is consistent with the
system being in chemical equilibrium. The extracted inverse slope parameters are 
the same for both $\xim$ and $\axi$ and are not strongly dependent with centrality. 
For central collisions this inverse slope parameter is higher than that from 
collisions at the SPS energy and shows a deviation from a collective flow pattern 
that may indicate an earlier freeze-out of multi-strange baryons.

\end{document}